# 65 NM CMOS SENSORS APPLIED TO MATHEMATICALLY EXACT COLORIMETRIC RECONSTRUCTION

C. Mayr[1], S. Henker[1], A. Krause, J.-U. Schlüßler, and R. Schüffny
Institute of Circuits and Systems, University of Technology Dresden
Helmholtzstr. 18, 01069 Dresden, Germany
{mayr,henker,krause,schlue,schueffn}@iee.et.tu-dresden.de
([1]The first two authors contributed equally to the research described in this manuscript.)

**ABSTRACT**

Extracting colorimetric image information from the spectral characteristics of image sensors is a key issue in accurate image acquisition. Technically feasible filter/sensor combinations usually do not replicate colorimetric responses with sufficient accuracy to be directly applicable to color representation. A variety of transformations have been proposed in the literature to compensate for this. However, most of those rely on heuristics and/or introduce a reconstruction dependent on the composition of the incoming illumination. In this work, we present a spectral reconstruction method that is independent of illumination and is derived in a mathematically strict way. It provides a deterministic method to arrive at a least mean squared error approximation of a target spectral characteristic from arbitrary sensor response curves. Further, we present a new CMOS sensor design in a standard digital 65nm CMOS technology. Novel circuit techniques are used to achieve performance comparable with much larger-sized specialized photo-CMOS processes. The sensor is utilized as testbed for the spectral reconstruction method.

## 1. Introduction

Colorimetric reconstruction starts out with a set of image sensors $j$ converting a range of incident illumination to a scalar sensor output signal. These sensors can be described by their spectral sensitivity functions $s_j(\lambda)$ which assign a wavelength–dependent conversion efficiency to the sensors. The integral of these conversion efficiencies multiplied by the spectral distribution of the incoming light $\varphi(\lambda)$ across a range of wavelengths ($a…b$) constitutes the scalar output $S_j$:

$$S_j = \int_a^b \varphi(\lambda) \cdot s_j(\lambda)\, d\lambda \qquad (1)$$

Different filter and sensor characteristics are employed to render these sensor responses close to the ones employed by the end user of an image acquisition, e.g. the human visual system (HVS) [1,2]. However, technically feasible sensor spectral characteristics tend to approximate these responses only to an unsatisfactory degree [2,3], making some kind of transformation between the actual sensor responses and the desired spectral characteristic of an ideal sensor necessary. Correspondingly, a target for spectral reconstruction is defined as the responses $A_i$ which a set of ideal sensor curves $a_i(\lambda)$ would produce from the incident illumination:

$$A_i = \int_a^b \varphi(\lambda) \cdot a_i(\lambda)\, d\lambda \qquad (2)$$

For colorimetric reconstruction, these sensors are usually taken to be the CIE XYZ curves (Fig. 1):

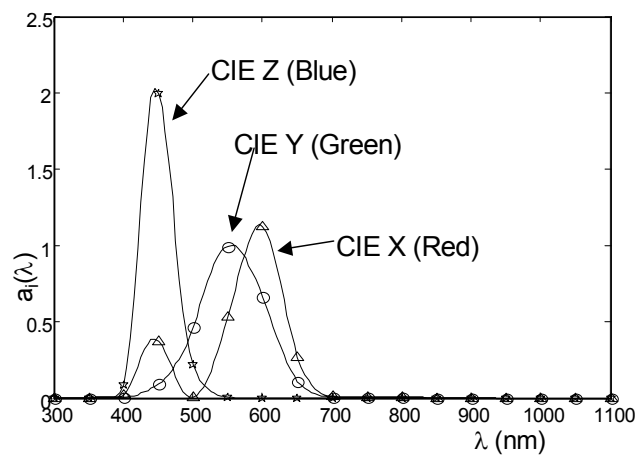

Fig. 1. CIE-XYZ curves, the y-coordinate is a dimensionless scalar of relative sensitivity.

These curves approximate the spectral sensitivities of the different color channels of the HVS [2]. In addition, the CIE responses have been widely accepted as device-independent exchange format [4,1], so deriving them from the sensor responses also aids in representing the image in the target application (printer, display, etc).

In the next section, we will give a short review of the reconstruction methods employed in current literature for this transformation between sensors and a target spectral characteristic. In section 2.2, a new method for least-squared-error matrix derivation for this transformation is derived. Section 3 introduces the new sensor IC, with 3.3 giving measurement results, while section 4 demonstrates the spectral reconstruction via sample images and reconstructed spectral response curves. The sensor design focuses on simplicity and inexpensiveness of design, using simple color filters and conventional digital CMOS processes. Several algorithmic and circuit techniques are detailed that compensate for the deficiencies incurred by this design strategy.

## 2. Colorimetric Reconstruction

### 2.1 Review of Current Colorimetric Methods

Illumination independent, global correction of non-colorimetric spectral responses of a color image sensor is only possible for linear combinations of the sensor spectral characteristics. As shown in [5], if any kind of nonlinear correction is attempted, a dependency on the sensor input spectrum $\varphi(\lambda)$ is introduced. Consequently, nonlinear correction methods tend to be linked to specific image stimuli [6, 7].

A method for color correction using this approach is the vector space approximation introduced in [6]. An estimation of the spectral distribution of the incident illumination is carried out from the sensor responses, which is then used to reconstruct the spectral space of an ideal observer. No analytical proof can be carried out if such an estimate will be successful in reconstructing the original light spectral distribution. For carefully selected, reduced experimental conditions, i.e. a select set of test images, this estimate usually leads to better spectral reconstruction results than simpler, linear estimates. This is caused by the fact that a reduction in test sets will lead to severely limited variations among the structure of the incident spectra, which enables reconstruction of the original colours via a classification approach. This method does not seem to fit more complex environmental conditions. For every data/image vector, all possible spectra would have to be searched to reconstruct the original colours. However, a decision as to which of the (theoretically infinite) spectral distributions has led to this specific sensor output is not possible without setting very strict boundary constraints.

A similar approach is carried out in [7], where the surface reflectance of image objects is estimated via a Bayesian metric. Again, better results are reported for this method compared to colorimetric reconstruction, but only for a specific set of surfaces encountered in the picture. Apart from the dependence on image composition, the authors admit that their algorithms' performance depends strongly on optimizing it for the 'scene illuminants likely to be encountered' (quote), thus it is also dependent on the ambient light composition.

Han et al. [4] see this illumination dependency as one of the major drawbacks of polynomial, i.e. nonlinear color correction methods and try to generalize it for arbitrary illumination conditions. They start out with a transformation matrix between sensors and ideal responses for a limited subset of image content. This is derived from adjusting color reconstruction for a range of 'target colors' at a specific white balance point, i.e. illumination composition. A method for adjusting this transformation matrix is then derived which takes into account variations in white balancing relative to the original characterization setup. Again, this approach is only valid for a subset of image content.

A linear combination of sensor channels is used in [2] to improve matching between target CIE curves and the CMOS sensor described in this publication. However, the linear combination matrix seems to be manually optimized, as evinced by the following quote from [2]: 'The green is reduced by the blue and the red to narrow the peak and the red channel is augmented with the green to shift the peak to a slightly lower wavelength'.

In [8], a matrix-based transformation is employed to map the 16 input channels of their multispectral image sensor to the 6 output channels of the display system, but no specifics are given how this matrix can be derived.

### 2.1 Proposed new Colorimetric Reconstruction

We attempt a reconstruction of the $m$ channels of the ideal (e.g. 3 channel CIE) observer characteristic in (2), denoted by the index $i$, via a linear combination of the $n$ sensor spectral characteristics (1), denoted by the index $j$:

$$A'_i = \int_a^b \varphi(\lambda) \cdot \sum_{j=1}^n \left[c_{ij} \cdot s_j(\lambda)\right] d\lambda \quad (3)$$

From these approximated color space coordinates, a new set of spectral response functions $a_i$ can be derived:

$$a'_i(\lambda) = \sum_{j=1}^n \left[c_{ij} \cdot s_j(\lambda)\right] \quad (4)$$

To achieve an accurate approximation of color space coordinates $A_i$ from the sensor space coordinates $S_i$, the deviation $F_i$ between $a_i$ and $a'_i$ across the spectral range $a…b$ has to be minimized:

$$F_i = \int_a^b \left[a_i(\lambda) - a'_i(\lambda)\right]^2 d\lambda \quad (5)$$

We extend the spectral reconstruction method presented in [5] with a strict derivation of the coefficients necessary for the Least Mean Squared Error (LMSE)-fit expressed in (5). Going back to the constituent elements of $a'_i$, (5) can be written as:

$$F_i = \int_a^b \left\{a_i(\lambda) - \left[c_{i1} \cdot s_1(\lambda) + … + c_{in} \cdot s_n(\lambda)\right]\right\}^2 d\lambda \quad (6)$$

The minimum of this function can be computed via the first partial derivative with respect to the coefficients $c_{ij}$:

$$\frac{\partial F_i}{\partial c_{ij}} = \int_a^b -2\left\{a_i(\lambda) - \left[c_{i1} \cdot s_1(\lambda) + … + c_{in} \cdot s_n(\lambda)\right]\right\} \cdot s_j(\lambda) d\lambda \quad (7)$$

The second partial derivative can be expressed as:

$$\frac{\partial^2 F_i}{\partial c_{ij}^2} = \int_a^b 2 \cdot s_j(\lambda) \cdot s_j(\lambda) d\lambda \quad (8)$$

Since the second derivative is always positive, the single zero crossing of (7) constitutes a minimum of the LSME-error function:

$$\int_a^b -2\{a_i(\lambda) - [c_{i1} \cdot s_1(\lambda) + \ldots + c_{in} \cdot s_n(\lambda)]\} \cdot s_j(\lambda) d\lambda = 0 \quad (9)$$

Equation reordering of (9) gives:

$$\int_a^b a_i(\lambda) \cdot s_j(\lambda) d\lambda = \int_a^b \{[c_{i1} \cdot s_1(\lambda) + \ldots + c_{in} \cdot s_n(\lambda)]\} \cdot s_j(\lambda) d\lambda \quad (10)$$

Since this derivation has to be carried out for all $j=1\ldots n$, the above equation can be written in vector format, with a column vector of spectral characteristics $\underline{s}=[s_1(\lambda)\ldots s_n(\lambda)]^T$ and a coefficient row vector $\underline{c_i}=[c_{i1}\ldots c_{in}]$:

$$\int_a^b a_i(\lambda) \cdot \underline{s}(\lambda)^T d\lambda = \underline{c_i} \cdot \int_a^b \underline{s}(\lambda) \cdot \underline{s}(\lambda)^T d\lambda, \quad (11)$$

with both sides constituting a row vector of the single correspondences expressed in (10). Equation (11) can again be extended across the whole range $i=1\ldots m$ of color space spectral response functions $a_i$ to form a set of vectorized equations:

$$\int_a^b a_1(\lambda) \cdot \underline{s}(\lambda)^T d\lambda = \underline{c_1} \cdot \int_a^b \underline{s}(\lambda) \cdot \underline{s}(\lambda)^T d\lambda$$
$$\vdots \quad (12)$$
$$\int_a^b a_m(\lambda) \cdot \underline{s}(\lambda)^T d\lambda = \underline{c_m} \cdot \int_a^b \underline{s}(\lambda) \cdot \underline{s}(\lambda)^T d\lambda$$

Once again defining a column vector $\underline{a}=[a_1(\lambda)\ldots a_m(\lambda)]^T$, (12) can be rewritten in matrix notation:

$$\int_a^b \underline{a}(\lambda) \cdot \underline{s}(\lambda)^T d\lambda = \Im \cdot \int_a^b \underline{s}(\lambda) \cdot \underline{s}(\lambda)^T d\lambda$$
$$\text{with} \quad \Im = \begin{bmatrix} \underline{c_1} \\ \vdots \\ \underline{c_m} \end{bmatrix} = \begin{bmatrix} c_{11} & \cdots & c_{1n} \\ \vdots & \ddots & \vdots \\ c_{m1} & \cdots & c_{mn} \end{bmatrix} \quad (13)$$

To derive an expression for the coefficient matrix $\Im$, both sides of (13) can be multiplied with the inverse of the integral on the right hand side, which gives:

$$\Im = \left[\int_a^b \underline{a}(\lambda) \cdot \underline{s}(\lambda)^T d\lambda\right] \cdot \left[\int_a^b \underline{s}(\lambda) \cdot \underline{s}(\lambda)^T d\lambda\right]^{-1} \quad (14)$$

To use this coefficient matrix with the sensor responses as expressed in (1), (3) can be rewritten in vector format (equation on the left side):

$$A'_i = \underline{c_i} \int_a^b \varphi(\lambda) \cdot \underline{s}(\lambda) d\lambda \quad resp. \quad \underline{A'} = \Im \cdot \underline{S} \quad (15)$$

Extending the vector format to the full matrix across all channels of the ideal observer characteristic and carrying out the sensor spectral integral yields the equation on the right side of (15). If the full expression for the coefficient matrix in (14) is inserted in the above equation and reordered, the following can be derived:

$$\underline{A'} = \int_a^b \underline{a}(\lambda) \cdot \left\{\left[\int_a^b \underline{s}(\lambda') \cdot \underline{s}(\lambda')^T d\lambda'\right]^{-1} \underline{s}(\lambda)\right\}^T d\lambda \cdot \underline{S} \quad (16)$$

Please note that the above expression has no limitation as to the number of input or target output channels it employs. In this paper, we target the three-channel CIE XYZ representation, but this derivation will provide a least-squared-error transformation for higher-dimensional image representations as well. For example, given a sufficient number of sensor channels, the six band representation of an image in [8] can also be targeted.

Using (16), conditions can be derived for this spectral reconstruction to be complete. The first case would be if $\underline{s}(\lambda)$ and $\underline{a}(\lambda)$ are linearly dependent from one another. This can also be determined from (13), where the reconstruction error is zero if $\underline{a}(\lambda)=\Im*\underline{s}(\lambda)$. Incidentally, (16) thus constitutes a formal proof of the Luther condition.

Alternatively, (16) can be viewed as a sensor response equation similar to (2), where the sensor constitutes the ideal observer characteristic $\underline{a}(\lambda)$ and the spectrum of the incident light is given by the following expression:

$$\varphi'(\lambda) = \left\{\left[\int_a^b \underline{s}(\lambda') \cdot \underline{s}(\lambda')^T d\lambda'\right]^{-1} \underline{s}(\lambda)\right\}^T \cdot \underline{S} \quad (17)$$

Basically, this is a back transformation of the sensor responses $\underline{S}$ via their individual spectral characteristics to the incident light used in (2). Of course, the incident light is characterized by a continuous spectral composition, representing an infinite-dimensional space. To reconstruct it from the single coordinates $S_j$ in the sensor response space, these coordinates must also encompass an infinite set.

Since every practical sensor system offers only a finite set of basis functions $\underline{s}(\lambda)$, in general no error-free reconstruction is possible. As discussed above, the reconstruction quality is thus dependent on the one hand on the quality of the basis functions, which can be estimated by how much these sensor spectral responses vary across the wavelength range and from each other. Large variations such as bandpass or bandstop behaviour (across different bands by different sensors) improve reconstruction [8,2]. On the other hand, if an imaging technology offers only sensor spectral responses that vary little across wavelength and from each other, comparable spectral reconstruction can still be achieved by increasing the number of input channels [5] (see also Fig. 7). An ideal sensor characteristic $\underline{a}(\lambda)$ can thus be achieved by using multiple sensor channels, even if those are individually defective in their spectral response. Multiple sensor channels can be realized e.g. by using a single photo element in a time multiplexed manner while controlling its spectral sensitivity through the depth of its depletion layer [3]. Another method for implementing multiple sensor characteristics is the vertical stacking of photo diodes, offered by CMOS or BiCMOS technologies [9, 2]. An example for the latter method is discussed in the following.

# 3. The 65nm Sensor Evaluation IC

## 3.1 IC Summary

Today's CMOS sensors are chiefly fabricated in 0.18μm or larger technologies [10,2,9] since smaller technologies exhibit several detrimental effects, such as increased leakage currents. However, the trend towards multi-megapixel cameras advocates design explorations in technologies well below 0.2μm. Additionally, image sensors in deep-submicron can take full advantage of the technology shrink for digital image restoration, balancing technological deficiencies and offering additional processing capabilities [11]. Consequently, the design discussed herein focuses on evaluation of a 65nm CMOS technology for active photosensor arrays and analog/mixed signal readout circuits. The application is aimed at embedded imagers for digital cameras in handhelds and alike, using standard digital processes for Systems on Chip. As far as the authors are aware, this is the first study of pixel sensors in sub-100nm CMOS technologies.

First measurement results of the IC have been presented in [12]. It contains a 128x116 pixel array of 6x6μm dual channel pixels with 46% fill factor (Fig. 4). Additional a 48x96 pixel array of smaller 3x3μm single channel pixels utilizing embedded transistor reset and 38% fill factor has also been included. Beside the pixel arrays, the 65nm evaluation chip also includes circuits for analog processing such as several amplifiers, a pipeline ADC and a ΣΔ-ADC and their associated digital controls:

All of the analog components can be accessed via an internal multiplexer/demultiplexer, to evaluate their performance separately or in various combined operation. For example, the different OTAs and rail2rail Opamps can be evaluated via external stimuli, used as analog readout amplifiers for the pixel voltages, or employed as input amplifiers for the AD converters. Both ADCs can be used for converting the clocked readout of the pixel matrices to 8 Bit digital signals, with the ΣΔ-ADC using an external decimation filter (not included in the current design). Due to the non-availability of digital standard cells at the time of design, the state machine and other digital configuration circuits where carried out as hand-crafted layout. The pixel matrices both carry out conventional correlated double sampling (CDS) to suppress fixed pattern noise. A summary of the IC features is given in the following:

- array of 116x128 double diode active pixels (6μm pixel size, 0.7x0.77mm$^2$ array size)
- array of 48x96 single diode active pixels (3μm pixel size for evaluation)
- analog readout data path (correlated double sampling column buffer, programmable gain amplifier, analog crossbar)
- 8bit pipelined cyclic AD-converter
- analog test circuits (ΣΔ-modulator, OTAs, Rail2Rail OpAmps)
- digital programmable bias circuits, digital scan path (programming, test, to reduce pin count)
- 14 external pins, 2.5V power supply, 1.08 mm$^2$ chip size

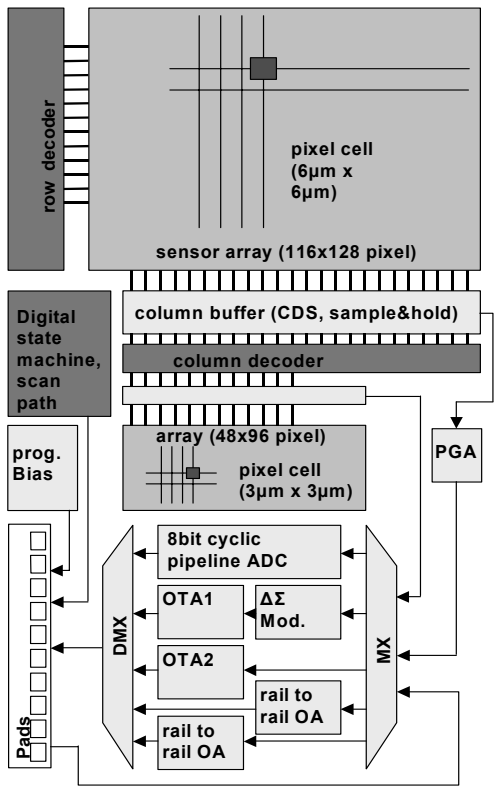

Fig. 2. Block diagram of the 65nm sensor evaluation IC

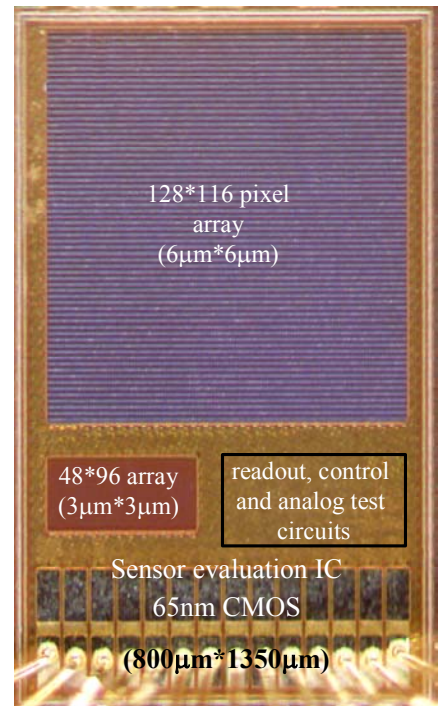

Fig. 3. Microphotography of the 65nm Sensor evaluation IC with main building blocks.

## 3.2 Pixel Description

For sensor operation we use linear light energy measurement by integrating a reverse biased junction diode capacitor. The measurement process consists of 3 phases: the resetting of the capacitor to a fixed voltage using a reset element, the integration of the light induced current over a distinct amount of time and the readout of the voltage difference on the capacitor using a readout element. The received light energy is then directly proportional to the voltage change on the capacitor. The low-active digital reset signal operating the transistor switches is applied at the RESET terminal (Fig. 4), readout to the bus is done via the two source followers. Two additional transistors at the right side of the circuit select the two photo diodes individually for readout, resulting in a 6T pixel cell [13].

The general problems with imager design in sub-100nm derive mainly from technology constraints [10,11]: The consequences of transistor scaling are increased leakage due to high doping concentration, steep implant profiles and increased oxide interface trap states caused by shallow trench isolation. Short channel effects significantly increase transistor off-current and some devices also show considerable gate leakage which further discharge integrating and sample&hold capacitors [11]. We aimed to counter this in part by using analog IO devices as transistors for the entire pixel cell. Analog IO devices show several advantages, esp. lower channel leakage and basically no gate leakage. In addition, specialized reset circuits and different reset schemes were implemented to reduce leakage. Specifically, in between resets, the reset levels at the terminals of the reset transistors are pulled to ground and the positive supply voltage for the well and the diffusion diode, respectively (Fig. 4). This causes the transistor channel leakage and gate tunneling current to be reduced and drawn towards the supply rails [14], away from the photo diodes.

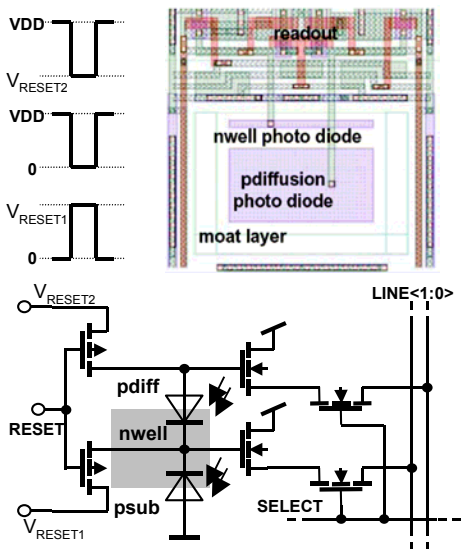

Fig. 4. Reset voltage scheme, layout and pixel cell reset and readout circuitry.

Further, we are using different configurations of junction diodes as photo element, focusing on well diodes, since the junction leakage of diffusion diodes is in general orders of magnitude higher because of their doping concentration [14]. Well diodes are not very common in pixel design due to their large interspace requirements. Our solution is to partially compensate this by applying moat layers, which prevent the doping implantation of the surrounding p-well. The resulting p-i-n diode structure shows reduced junction leakage and an enhanced depletion region as well as smaller capacitance and thus improved sensitivity. However, due to the implantation depth, a well diode shows higher sensitivity especially in the infrared region compared to shallow diffusion diodes [9]. An additional diffusion diode was also included in the pixel cell to form vertically stacked photo diodes with distinguishable spectral characteristics.

These pixel structures require separate reset and readout for the different diodes, but provide multiple output signals sampled on identical spatial and temporal coordinates, which are usable for the spectral reconstruction as discussed in section 2.

## 3.2 Spectral Modeling and Measurement

One of the main drawbacks of using a standard digital technology without any postprocessing is the interference effects of the metal stack above the photo diodes. The metal itself can be kept from covering the photo diode, but the silicon dioxde layers supporting the metal cover the whole IC, providing multiple paths for incident light to reach the surface.

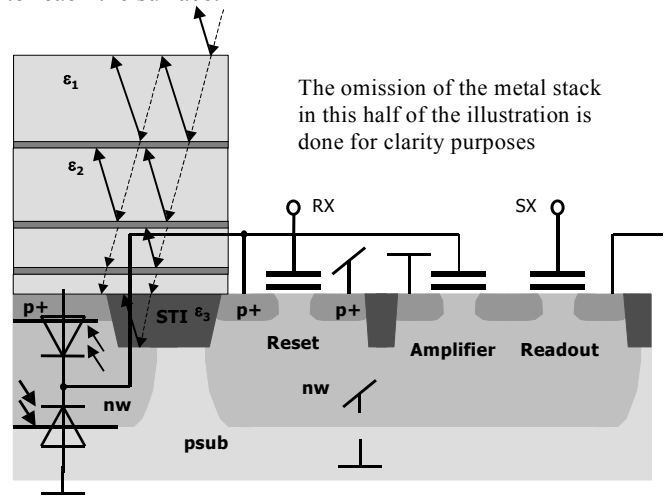

Fig. 5. Illustration of the reflective nature of the metal stack, including part of the photo diodes and readout circuitry. For an actual scanning microscope image of the metal stack atop a CMOS photodetector, see Fig. 1 in [15]

These multiple paths are especially pronounced in a technology using copper, since copper needs additional dielectric layers preventing diffusion of the copper in the silicon dioxide. These dielectric layers are highly reflective for optical rays. Since the dimensions of the metal stack are in the wavelength range of visible light, wave-optical effects produce strongly varying transmission characteristics across the spectral characteristic of the photo diodes [9].

Usually, this effect is avoided through specialized processing options, e.g. the sensor described in [2] removes the complete metal stack (including the support layers) above the photo diode and covers the sides of the resulting pit with a non-reflective layer. Another option is to use completely non-standard technologies, which directly integrate photo sensors on some substrate [3]. Even such a technology has to battle with interference effects, which Knipp et al. [3] mention but do not take into account in their modelling of the sensor spectral characteristics. Additionally, the technological limit of such specialized processes usually extends only as far as including a readout transistor with the pixel cell, no further active elements can be integrated with the pixel cell. Consequently, the various image processing functions easily included in a CMOS technology must in this case be carried out externally [11]. Both of the above approaches also add greatly to the cost of the sensor manufacturing process compared to a run-of-the-mill CMOS technology.

The approach taken in this design is to model the interference effects to predict sensor characteristics for a given technology, so the designer can simply choose the one with the least detrimental effect on his targeted spectral range. In [15], a wave optical approach is described to model light intensity dependent on the incident angle. We used a similar method to derive the spectral transmission characteristic of the 65nm copper metal stack as a function of wavelength. The result of this modeling is illustrated in the uppermost curve in Fig. 6, corresponding to the graph annotation at the right side:

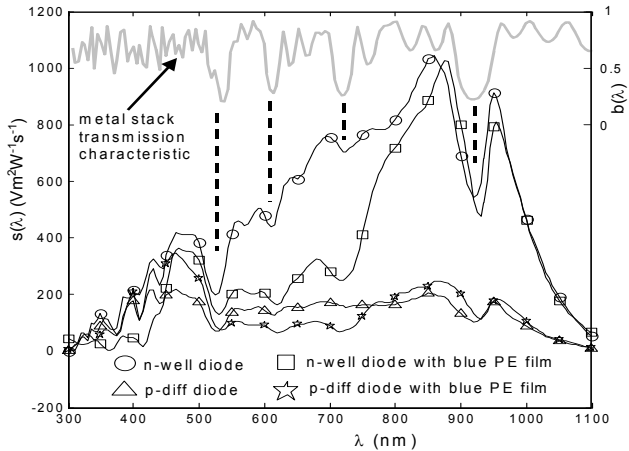

Fig. 6. Modelled metal stack transmission characteristic (uppermost curve and right graph annotation) and measured spectral response curves of various diode and filter configurations.

The annotation is given as a dimensionless metal stack transmission coefficient, ranging from total blockage (zero) to unhindered transmission (one). The minima correspond well with the ones observed in the spectral characteristics measured on the actual IC (dashed lines), confirming the validity of this approach. To derive a complete model of the sensor spectral responses, the transmission curve must be overlaid with the response curves of the actual sensors. The depth dependent photo charge generation and optical efficiency of various diode structures is well characterized and modeled [9,3], so this should present no problem.

The lower four curves in the Fig. 6 present measurements of the sensor spectral characteristics of the two diodes described in Fig. 4. Measurements have been carried out with a monochromator employing a bandwidth of 5nm. Two of the curves represent the spectral response when the IC is directly connected to the monochromator (as denoted by the figure caption). For the other two, an ordinary blue plastic film has been inserted in the optical path, to provide the spectral reconstruction carried out in the following sections with additional input channels. Especially the well diodes show a strong sensitivity to infrared wavelength. This can lead to image colour artefacts and is usually controlled through a filter [2]. In contrast, we will show in the next section that the spectral reconstruction can also be employed to carry out this infrared suppression. Further characterization of other optical parameters of the photo diodes leads to the following results:

| Feature | Unit | 65nm, nwell with moat | 65nm, nwell/pdiff with moat | 180nm Reference [10] |
|---|---|---|---|---|
| pixel size | μm$^2$ | 6x6 | 6x6 | 3.5x3.5 |
| integrating capacitor | C / fF | 11 | 15.3 | 4.5 |
| **dark** current sensitivity | $S_{dark}$/mV/s | 50 | 81 | 6.29 |
| saturation | Nsat / e- | 82397 | 110342 | 22472 |
| reset noise[1] | Nrn / e- | 42 | 57 | 20 |
| dark current noise | Ndn / e- | 69 | 153 | 4 |
| noise floor | Nn / e- | 81 | 163 | 21 |
| SNR[2] | dB | 60.2 | 54.8 | 61 |
| [1] reset noise is assumed to be 1* sqrt(kT /C), not yet measured, (Reference [10]: 0.75* sqrt(kT /C) | | | | |
| [2] based on measurements of dark current and reset noise estimation | | | | |

Table 1 Comparison of 65nm Pixels with IBIS-Sensor

Measurement results thus prove that important imager parameters like sensitivity, dark current, and resolution in a standard deep-submicron CMOS technology are comparable with state of the art CMOS-imagers using special photo process options (as compared to reference [10] in Table 1, see also the performance figures reported in [2]). Thus, the corrective measures described in section 3.2 seem to be effective in extending the range of CMOS technologies usable for image sensors well below 100nm, employing conventional CMOS processes without any specialized post-processing options.

## 4. Colorimetric Results

### 3.1 Spectral Correction

The four sensor spectral responses of Fig. 6 have been used as input for the spectral reconstruction outlined in section 2.2, with the CIE curves of Fig. 1 as colorimetric reconstruction target. Fig. 7a illustrates the results obtained from this reconstruction. The blue CIE spectral characteristic is executed reasonably well, on a par with the B-channel of a Bayer-Mosaic-RGB camera reported in [1].

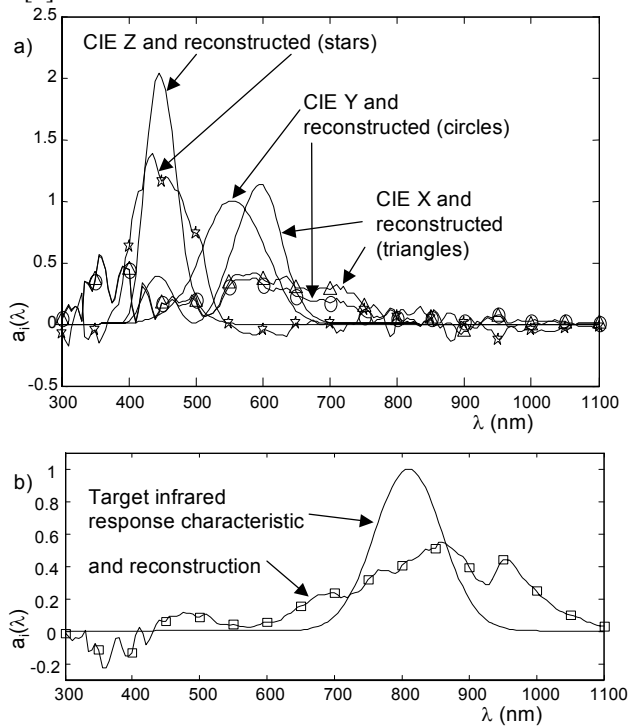

Fig. 7. (a) CIE-XYZ targets and actual reconstruction from 65nm IC, (b) additional infrared channel reconstructed from the sensor channels.

The red and green channels are not very well separated due to the insufficient variation between single sensor responses in this wavelength range as shown in Fig. 6. But even the small difference in the slopes of the reconstructed red and green CIE curves from 600 to 750 nm is sufficient for a colour rendition comparable to a conventional RGB-filter (see Fig. 8), while the infrared response is much better controlled in all three channels compared to RGB, whose infrared response extends well beyond 800nm [1]. The suppression of the infrared response possible in a setup using four sensor channels as compared to conventional RGB is also demonstrated by the fact that we can offer an additional infrared target spectral characteristic (Fig. 7b, target spectral response and actual reconstruction). This channel might e.g. be applicable to specialized heat-based image analysis functions.

In addition, the RGB-filter produces more colour moiré, since it has to split its colour acquisition into four spatially separate coordinates [1,3], while the above spectral response would only need two, i.e. a filter mask with blue and empty squares in a checkerboard layout.

If more than four sensor characteristics are available, the potential for CIE (or other target spectral responses) reconstruction is greatly improved [5].

### 3.1 Sensor Sample Image

In [5], a sensor simulation toolkit has been introduced, which can be used to model various aspects of an image acquisition and spectral processing chain. It includes a model for a complete optical path with its ambient illumination, image object and photosensor device. Using this model, a sample image has been processed with a (simulated) sensor exhibiting RGB spectral responses similar to the ones reported in [1], and with the reconstructed color channels of Fig. 7. The image has been overlaid with an infrared disturbance to judge colour rendition for both sensors in the presence of infrared illumination. As seen below, the RGB sensor tends towards a green distortion of the input image, while the colour channels reconstructed from the 65nm IC amplify the blue portion.

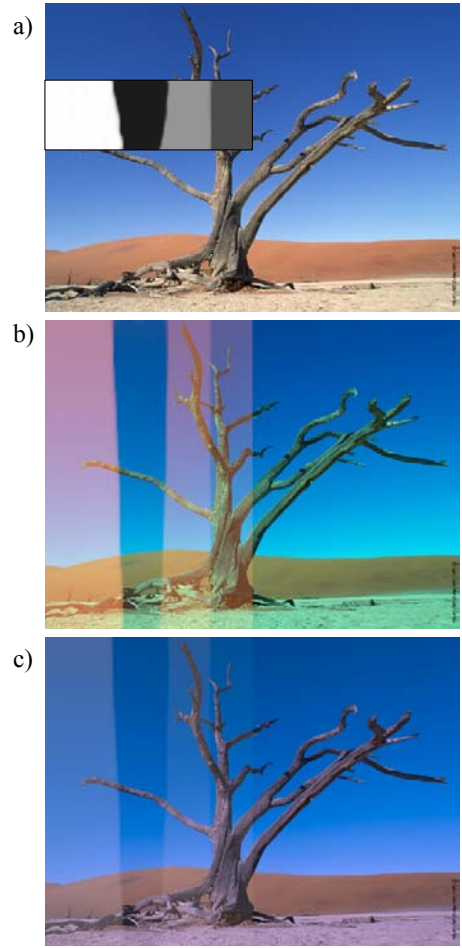

Fig. 8. (Simulated) image acquisition results, (a) original image plus infrared distortion, (b) generic RGB sensor, (c) CIE channels reconstructed from 65nm spectral responses.

Visually, both artefacts seem to be about equal in their distortion of the original image. As observed for the spectrally reconstructed colour channels in the last section, the infrared response is much better controlled for the colour channels reconstructed from the 65nm IC.

The colour reproduction as shown above (at the resolution offered by the pixel matrix) can be reached with the current 65 nm IC in several different ways. The most conventional one would be with the above-mentioned checkerboard blue/empty structured colour filter. Secondly, the IC could be used in a time interpolated manner, with one image taken of a target object with the sensor alone, and a second image taken while the same blue PE plastic film is inserted in its optical path as the one used for measuring the spectral characteristics in Fig. 6. The most sophisticated way would be with two samples of the 65nm IC aligned on the same target, where one of them employs the blue PE film in the optical path. This would realize four sensor channels taken of the same object at the full temporal and spatial resolution offered by the sensor. We aim to present such a demonstrator for both the technology and the spectral reconstruction method at the conference, offering RGB capability with a very inexpensive setup.

## 5. Conclusion

As put forward in the review in section 2.1, current colorimetric reproduction relies either on properties of the incident illumination and thus can be used only for restricted test cases [4,6,7]. In other cases, color filters are used in such a way that the sensor spectral responses already have very good bandpass characteristics. For multichannel bandpass sensors, colorimetric response curves can be used as envelopes, where bandpass sensor responses are simply scaled and added to construct a piecewise approximation of the target curve [8]. Alternatively, simple heuristics can be used for a limited number of input channels to enhance their individual correlation with target CIE curves [2]. We put forward a method that is more general than any of the above. First, it does not rely on illumination properties, second, an arbitrary number of input and target spectral responses can be used. Third, spectral responses of the input channels do not have to exhibit any bandpass or other special behaviour. Moreover, this method can be strictly mathematically derived and applied via simple linear matrix computations. Matrix coefficients can be precomputed based on spectral response measurements of the image sensors employed. In the second part of the paper, the first image sensor in a 65nm CMOS technology is presented (to the best of our knowledge). Sensor measurements confirm the validity of novel circuit techniques employed to enhance the sensor-related performance of this technology on a par with specialized, much larger CMOS technologies. Practical examples of the spectral reconstruction method using the 65nm IC confirm its applicability.

## Acknowledgements

The authors would like to gratefully acknowledge the financial support by the Infineon AG, Munich subsidiary, division corporate research.